\documentclass[structabstract]{aa}  

\usepackage{graphicx}
\usepackage{natbib}
\bibpunct{(}{)}{;}{a}{}{,} 
\usepackage{amsmath}
\usepackage{url}
\usepackage{txfonts}
\usepackage{multirow}
\usepackage{setspace}
\pdfoutput=1

\begin{document}

\title{The mid-infrared extinction law in the darkest cores of the
  Pipe Nebula\thanks{Based on observations made with ESO Telescopes at
    the La Silla Paranal Observatory under programme IDs 62.I-0866,
    65.I-0083, 67.C-0514, and 069.C-0426.}}

\author{Joana Ascenso\inst1 \and Charles J. Lada\inst{2} \and Jo\~ao
  Alves\inst{3} \and Carlos G. Rom\'an-Z\'u\~niga\inst{4} \and Marco
  Lombardi\inst{5}}

\institute{European Southern Observatory, Karl-Schwarzschild-Str. 2,
  85748 Garching bei M\"unchen, Germany \and Harvard-Smithsonian
  Center for Astrophysics, 60 Garden Street, Cambridge, MA 02138, USA
  \and University of Vienna, T\"urkenschanzstrasse 17, 1180 Vienna,
  Austria \and Instituto de Astronom\'ia, Unidad Acad\'emica de
  Ensenada, Universidad Aut\'onoma de M\'exico. Ensenada 22860
  M\'exico \and University of Milan, Department of Physics, via
  Celoria 16, 20133 Milan, Italy}

\date{}

\abstract
{The properties of dust grains, in particular their size distribution,
  are expected to differ from the interstellar medium to the
  high-density regions within molecular clouds.}
{We measure the mid-infrared extinction law produced by dense material
  in molecular cloud cores. Since the extinction at these wavelengths
  is caused by dust, the extinction law in cores should depart from
  that found in low-density environments if the dust grains have
  different properties.}
{We use the unbiased LINES method to measure the slope of the
  reddening vectors in color-color diagrams. We derive the
  mid-infrared extinction law toward the dense cores B59 and FeSt
  1-457 in the Pipe Nebula over a range of visual extinction between
  10 and 50 magnitudes, using a combination of Spitzer/IRAC, and ESO
  NTT/VLT data.}
{The mid-infrared extinction law in both cores departs significantly
  from a power-law between 3.6 and 8 $\mu$m, suggesting that these
  cores contain dust with a considerable fraction of large dust
  grains. We find no evidence for a dependence of the extinction law
  with column density up to 50 magnitudes of visual extinction in
  these cores, and no evidence for a variation between our result and
  those for other clouds at lower column densities reported elsewhere
  in the literature. This suggests that either large grains are
  present even in low column density regions, or that the existing
  dust models need to be revised at mid-infrared wavelengths. We find
  a small but significant difference in the extinction law of the two
  cores, that we tentatively associate with the onset of star
  formation in B59.}
{}


\titlerunning{The mid-infrared extinction law in the darkest cores of the Pipe Nebula}
\authorrunning{J. Ascenso et al.}

\maketitle

\section{Introduction}
\label{sec:introduction}

Dust plays a crucial role in many of the chemical and physical
processes that take place in the interstellar medium. It plays an
especially important role in molecular clouds and in the formation of
stars. At low densities, dust grains are the catalysts for the
formation of molecules that then cool the clouds radiatively
\citep[e.g., ][]{Whitworth98, Draine11}. At high densities the
molecules begin to deplete from the gas phase and dust becomes the
dominant coolant, lowering the gas temperature enough that
fragmentation of cores into stars may occur \citep{Whitworth98,
  Larson05}. Yet, the properties of the dust grains, including their
composition and size distribution, and how they evolve in different
environments are not accurately known.

As the density increases, dust grains are believed to grow by
coalescence through grain-grain inelastic collisions and by the
accretion of ice mantles, as molecules freeze out on to their surfaces
\citep[e.g.,][]{Whittet88,Ossenkopf93}. The ice mantles are typically
very thin \citep{Draine85} and so do not increase the grain size
significantly by themselves, but they promote further growth by making
the grains ``stickier'' in collisions \citep[e.g.,][]{Ormel09}. The
size distribution of dust grains is therefore expected to vary from
the low-density interstellar medium to the cold, dense environment of
cores as the small grains and PAHs (polycyclic aromatic hydrocarbons)
disappear and populate the larger grain-size side of the distribution.

The impact of grain growth on star formation is not clear, but there
are indications that it may play a part. \citet{Larson05} suggests
that the grain size distribution is not relevant to the control of
temperature at the highest densities ($>10^{-18}$ g cm$^{-3}$), but
that it can be important at intermediate densities where it can still
affect the collision rate between the gas and dust. This defines the
density at which gas and dust become thermally coupled, which
according to the author ultimately determines the characteristic
stellar mass. \citet{Omukai00}, on the other hand, suggests that
different grain size distributions can produce different minimum
temperatures, again influencing the fragmentation properties of the
cloud, although, according to \citet{Larson05}, their models produce
unrealistically low temperatures considering the radiative heating by
ambient far-infrared radiation.

The extinction law is often used as a proxy for the size distribution
and composition of dust grains, traditionally in the optical and
ultra-violet \citep[e.g.,][]{Savage79,Draine11} and more recently also
in the near- and mid-infrared. The optical and ultraviolet have been
successfully used for decades to study the low-density interstellar
medium, revealing an extinction law that can be described as a
function of a single parameter, the total-to-selective extinction
ratio, $R_V=A_V/E(B-V)$ \citep{Cardelli89,Olofsson10}. In the
near-infrared (NIR) the extinction law is considered invariant and
well described by a power law ($A_\lambda \propto \lambda^{-\alpha}$)
of index $\alpha$ in the range $1.6-1.85$
\citep{Koornneef82,Whittet88,Cardelli89,Martin:1990zr,Clayton88,Becklin78,Landini84,Rosenthal00,Nishiyama:2006ve,Olofsson10},
and in the mid-infrared (MIR, 3-8 $\mu$m in this paper), the
wavelength range of interest for the present study, the extinction law
is not yet well-characterized or understood. \citet{Steinacker10} have
proposed another tool to probe the grain size distribution of dust
using scattering of light rather than absorption. They used the
``coreshine'' method to infer the presence of large (up to 1 $\mu$m)
grains in dense molecular cores.

In this paper we compare the infrared extinction law in two dense
cores of the Pipe Nebula: \object{FeSt 1-457}, which is quiescent, and
\object{B59}, which contains a small young cluster
\citep{Forbrich09,Covey10}. The Pipe Nebula is one of the closest
molecular clouds to the Sun \citep[$d = 130$ pc, ][]{Lombardi:2006gf}
and has been studied extensively over the few past years. The first
systematic study of the cloud belongs to \citet{Onishi99}, who present
CO observations of a large part of the cloud and estimate the total
$^{12}$CO mass to be of the order of $10^4$ M$_\odot$.
\citet{Lombardi:2006gf} refine the distance to the cloud and present a
high-resolution extinction map based on 2MASS data covering
$8^{\circ}\times6^{\circ}$, later improved by \citet{Roman-Zuniga10}
using deep infrared data. \citet{Alves07} identify and characterize
159 cores within the Pipe based on the extinction map, ranging from
3.0 to 24.3 visual magnitudes of extinction, and with corresponding
masses between 0.5 to 28 M$_\odot$, and use this information to derive
the Pipe's dense core mass function. \citet{Muench07} establish most
of these cores as actual dense structures, rather than the
superposition of less dense material, using observations of the
$J=1-0$ transition of C$^{18}$O, further analyzing the cores's
kinematical structure. \citet{Rathborne08} further validate the
association of the cores with dense gas using molecular line
observations of NH$_3$ (1,1), NH$_3$ (2,2), CCS (2$_1$-1$_0$), and
HC$_5$N (9,8). Using the C$^{18}$O radio data and the data from the
extinction map, \citet{Rathborne09} produce a refined sample of cores
and confirm the previously found core mass function. \citet{Frau10}
and \citet{Frau12} study the size, density and chemistry of a few
cores in the Pipe using radio data, proposing an evolutionary scenario
for the chemistry in the cloud. The magnetic fields of the Pipe were
studied by \citet{Alves:2008aa} using optical polarimetry. They divide
the cloud in three zones according to the properties of the magnetic
field: B59, the densest core in the cloud, the long ``stem'', and
finally the ``bowl'', with the strongest magnetic field (15\%
polarization). They estimate the mass-to-flux ratio to be
approximately super-critical toward B59 and sub-critical inside the
bowl. Finally, \citet{Peretto12} and \citet{Gritschneder12} address
the formation of the cloud based on {\it Herschel} data and analytical
calculations. The extinction law in the Pipe was first studied in the
NIR by \citet{Lombardi06}, who find an average normal extinction law
for the whole cloud using 2MASS data. \citet{Roman-Zuniga07} study the
extinction law of the B59 core in particular and find it to be flat in
the $3.6-8.0~\mu$m regime, while confirming a normal law in the NIR.

In this paper we will readdress the extinction law of B59 using the
LINES method presented in \citet[][hereafter referred to as Paper
I]{Ascenso12}, and determine the extinction law of another dense core
in the Pipe, FeSt 1-457. This core is located in the opposite extreme
of the nebula from B59, and has no signs of active star
formation. \citet{Aguti:2007aa} suggest the core is gravitationally
bound in a quasi-stable state near hydrostatic equilibrium, while the
outer layers of the core are proposed to be oscillating.

Studying the extinction law in the context of the dust properties in
two cores of the same cloud has the clear advantage that the chemical
composition and in general the environment and conditions external to
the cloud must be the same. The differences between the two cores can
therefore be isolated and studied in detail.

In Section \ref{sec:observations} we present the data used for this
study, and in Section \ref{sec:extinction-laws} we derive the
extinction laws. In Section \ref{sec:discussion} we discuss the
results in the context of previous studies and models, and we present
our conclusions in Section \ref{sec:conclusions}.

\section{Observations}
\label{sec:observations}

This section describes the data used to derive the extinction law in
B59 (J2000, 17h11m21s, $-27^\circ26'10''$) and FeSt 1-457 (J2000,
17h35m47s $-25^\circ33'09''$), in the Pipe Nebula molecular cloud. The
data used here will be made available in the context of a forthcoming
paper.

\subsection{B59}
\label{sec:b59-data}

We used the dataset for B59 of \citet{Roman-Zuniga07}. The
near-infrared data were obtained with ESO NTT/SOFI and ESO VLT/ISAAC
in the $H$ (1.6 $\mu$m) and $K_s$ (2.2 $\mu$m) bands. The $J$ band
(1.2 $\mu$m) was only partially observed, and the catalog was
therefore complemented with 2MASS\footnote{The Two Micron All Sky
  Survey is a joint project of the University of Massachusetts and the
  Infrared Processing and Analysis Center/California Institute of
  Technology, funded by the National Aeronautics and Space
  Administration and the National Science Foundation.} data for better
coverage. This added 162 sources with respect to the original catalog
of \citet{Roman-Zuniga07}. Unfortunately, because 2MASS is too
shallow, there are still no sources with $J$ band information in the
center of the core.

The mid-infrared ({\it Spitzer}/IRAC) dataset was the exact same as in
\citet{Roman-Zuniga07}: the publicly available catalog, part of the
Cores to Disks Legacy Project \citep{Evans03}.

\subsection{FeSt 1-457}
\label{sec:fest-data}

\subsubsection{NIR data}
\label{sec:fest-nir}

The main observations for FeSt 1-457 were made with the ISAAC and the
SOFI near-infrared imagers, mounted, respectively, at the 8.2 m UT3
unit of the ESO/VLT, and at the 3.5 m ESO/NTT. The FeSt 1-457
observing runs were completed in March 1999 (SOFI, program ID
62.I-0866) and July 2000 (ISAAC, program ID 65.I-0083). All the data
are currently available in the ESO archives. SOFI has a field of view
(FOV) of $5'\times5'$ with plate scale of 0.288 arcsec pixel$^{-1}$,
almost seven times the angular resolution of 2MASS, and higher
sensitivity: achieved limits are estimated to be about 4-5 mag deeper
than 2MASS in each band.  Such characteristics can resolve the densely
crowded Galactic Bulge field in the background of the Pipe Nebula and
penetrate in regions with up to about 50 mag of visual extinction. The
field of view of SOFI is wide enough to completely contain some of the
larger dense regions in the Pipe Nebula (R $\sim$ 0.1–0.15 pc), with
the exception of the central core in Barnard 59. The ISAAC
observations were intended to fully resolve the centers of the most
dense and obscured regions.  These observations have a resolution of
0.144 arcsec pixel$^{-1}$ with a field of view a quarter of the size
of that of SOFI.

\subsubsection{MIR data}
\label{sec:fest-mir}

The {\it Spitzer}/IRAC\footnote{This work is based in part on
  observations made with the Spitzer Space Telescope, which is
  operated by the Jet Propulsion Laboratory, California Institute of
  Technology under a contract with NASA.} dataset for FeSt 1-457 is
from the Cycle I GTO observations and from the Cycle II GO Pipe Cloud
Survey. The data were available in the archive as processed by the
{\it Spitzer Science Center} team using pipeline version S18.7 that
already includes artifact correction. Each field was observed with
long (10.4s) and short (0.4s) exposure times. The source detection and
photometry were done using the
MOPEX\footnote{\url{http://irsa.ipac.caltech.edu/data/SPITZER/docs/dataanalysistools/tools/mopex/}}
pipeline software APEX multiframe on the artifact-corrected
images. The main difficulty in this step was the extreme crowding
caused by the position of the Pipe Nebula against the Galactic
Bulge. The crowding was severe in all fields except toward the very
center of the core where the extinction eliminates most of the
(background) sources. Apart from the technical problems that arise
when the software attempts to fit several PRFs (Point Response
Functions) to many ``clustered'' sources at once, the crowding
ultimately produced uncertainties in the photometry of sources with
bright neighbors. Unlike the typical photometric error, estimated by
the pipeline during the detection and fitting procedure, these errors
are difficult to quantify on an individual basis. Since the crowding
is a function of wavelength, Channel 1 ($3.6~\mu$m) being the most
crowded due to its higher sensitivity, these uncertainties will be
larger for the shorter wavelength bands. The source extraction and
photometry were done independently for the long and short exposures.

Only the objects with the best quality photometry were selected. The
quality criteria were a relative flux error below 10\%, and a $\chi^2$
of the PRF fit per degree of freedom below 25, which ensured a clean
sample. After the error cuts, the long and short exposure catalogs for
the same channel were merged such that, for the stars common to both,
those with the smaller SNR were discarded. The final catalogs for the
different channels containing only the sources obeying these criteria
were then merged with the NIR data.

\subsection{Control field}
\label{sec:control-field}

No dedicated control field was taken for this study. Instead we used
an extinction-free nearby field close to the core B72, the same that
was used by \citet{Roman-Zuniga07} in their work. For this field the
near-IR observations were from the same ESO/SOFI survey as described
in section \ref{sec:fest-nir}, and the mid-infrared data were from the
Spitzer/IRAC catalog of the Cores to Disk (c2d) Legacy project
\citep{Evans03}. The measured photometric dispersion for background
sources is comparable to the uncertainties in the $(\lambda - K_s)$
colors. This indicates that the intrinsic color dispersion in the
Galactic Bulge background field of the Pipe Nebula is small enough
that our photometric accuracy cannot resolve it. Because these data
were not meant as control fields for the present study, they are
shallower than the science observations.

\section{The extinction law in B59 and FeSt}
\label{sec:extinction-laws}

\subsection{LINES determination of $A_\lambda/A_\mathit{K_s}$}
\label{sec:lines}

\begin{figure}
  \resizebox{\hsize}{!}{\includegraphics{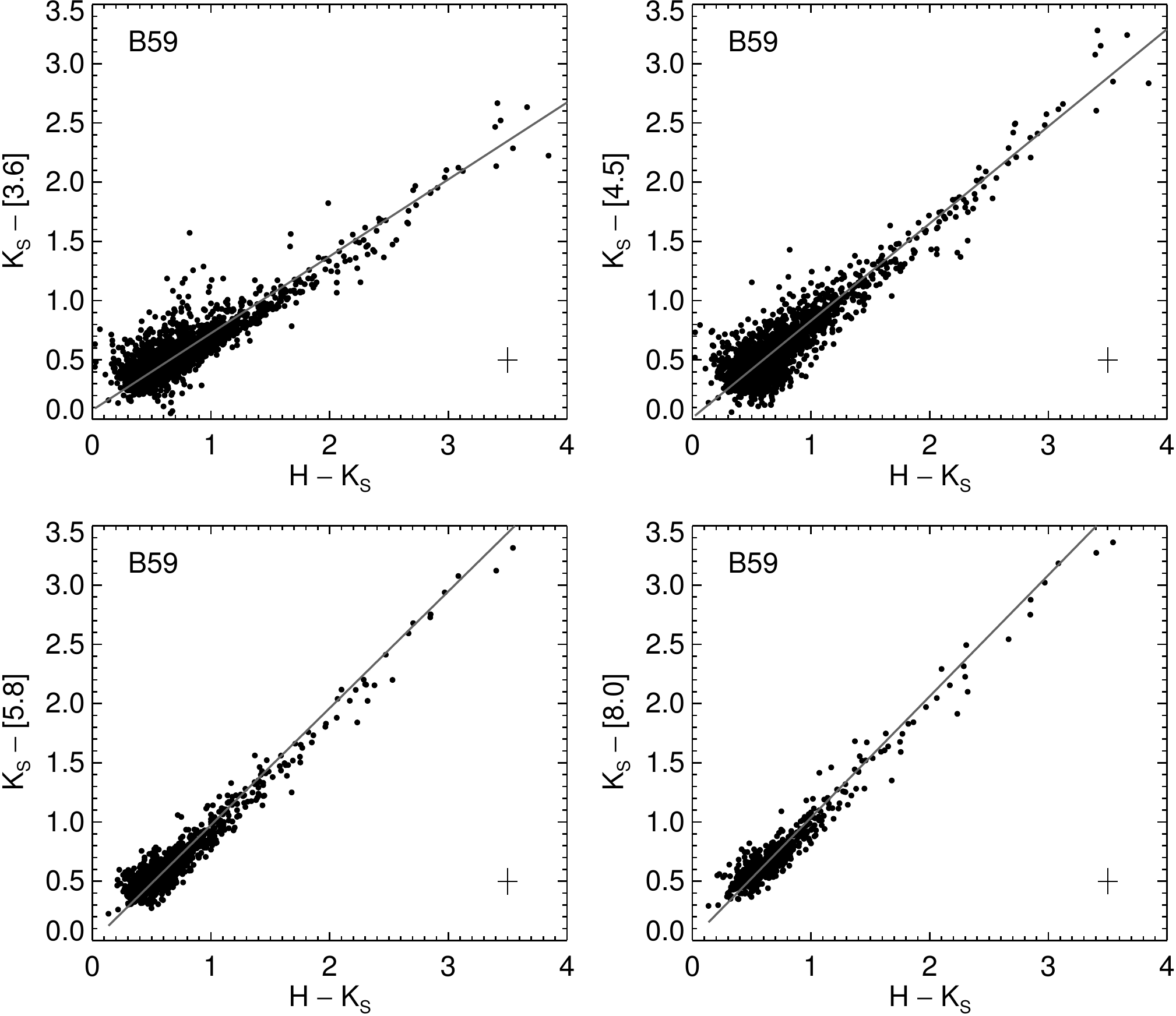}}
\caption{$(K_s-\lambda)$ {\it vs.} $(H-K_s)$ color-color diagrams for
  B59, for $\lambda=3.6$~$\mu$m ({\it top left}), $\lambda=4.5$~$\mu$m
  ({\it top right}), $\lambda=5.8$~$\mu$m ({\it bottom left}), and
  $\lambda=8.0$~$\mu$m ({\it bottom right}). The {\it solid lines}
  illustrate the reddening vectors for each band: the slopes are the
  LINES estimates, and the intercepta are such that the lines pass
  through the median of the distributions.}
\label{fig:ccd_b59}
\end{figure}

\begin{figure}
  \resizebox{\hsize}{!}{\includegraphics[width=12cm]{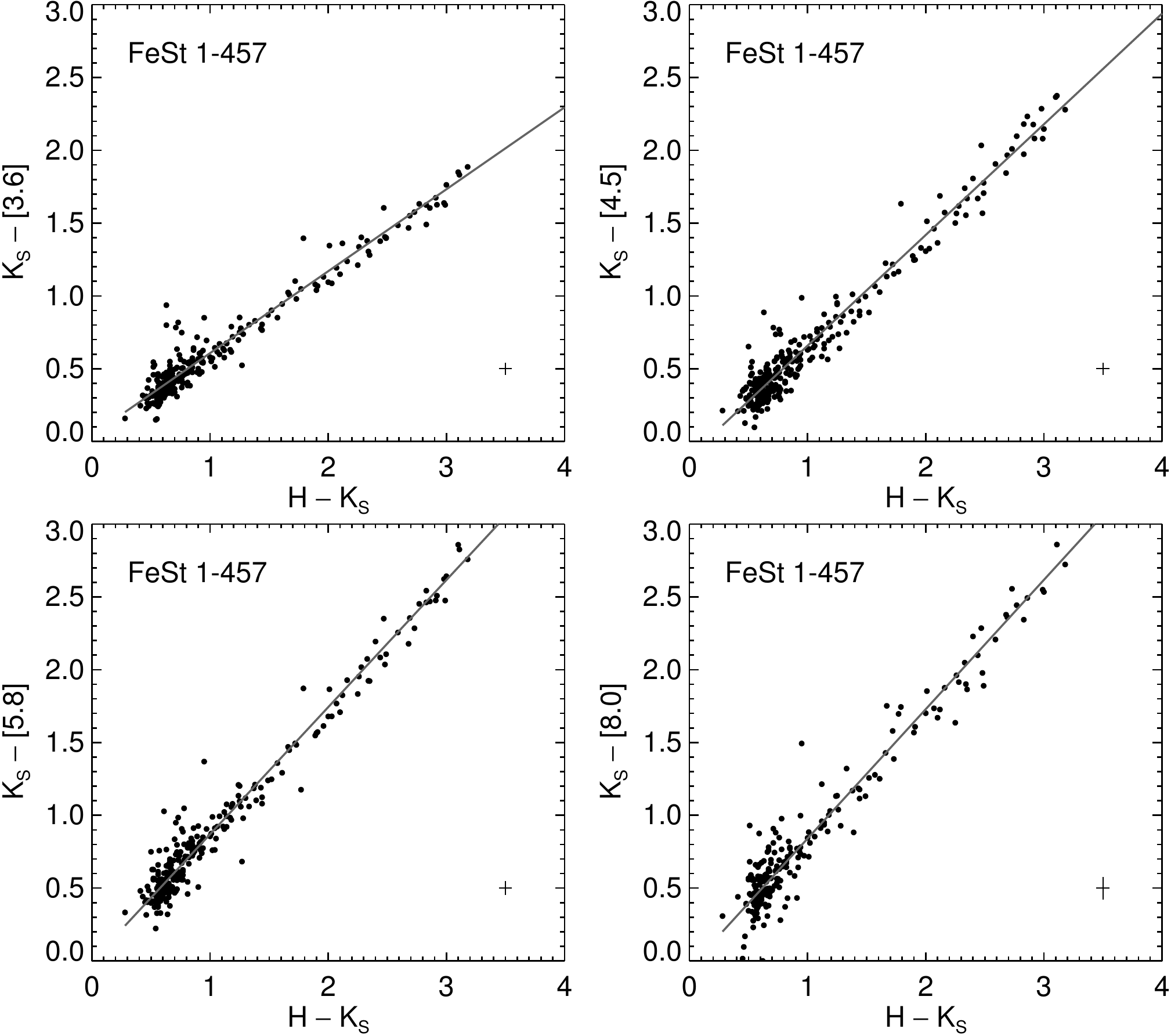}}
\caption{Same as Figure \ref{fig:ccd_b59} but for FeSt.}
\label{fig:ccd_fest}
\end{figure}

The data described above was used to construct $(K_s-\lambda)$ {\it
  vs.} $(H-K_s)$ color-color diagrams, where $\lambda$ is 3.6, 4.5,
5.0 and 8.0 $\mu$m for the IRAC bands. Since the observed stars are
mostly giants behind the FeSt and B59 cores, their position in the
color-color diagrams is dictated by the loci of their intrinsic
colors, by their photometric error, that scatters the points randomly
in the diagram, and by the extinction caused by the cores, that moves
the points along the reddening vector. The latter is the dominant
effect, as can be seen in Figures \ref{fig:ccd_b59} and
\ref{fig:ccd_fest}, and it is the slope $\beta$ of these vectors in
the different color-color diagrams that we wish to measure in order to
derive the extinction laws. LINES ({\bf Lin}ear regression with {\bf
  E}rrors and {\bf S}catter) has been validated in Paper I as the best
and most unbiased method to do this.

The LINES method determines the slope of a linear distribution of
points with errors in both directions, possibly correlated, and with
an intrinsic distribution that may not itself be linear. In the case
of the reddening vector in a color-color diagram, the intrinsic
scatter is the locus of the stellar population that is being reddened
by the cloud, and we wish to disentangle this scatter from the pure
contribution of the extinction. The intrinsic distribution of the
points is sampled by a control field expected to sample the same
stellar population. As described in section \ref{sec:control-field}
our control field was observed with a different setup from the science
field, producing a shallower dataset. However, since the background
population of the Pipe Nebula is dominated by giant stars which have a
very narrow range in infrared colors, the control field, albeit
shallow, represents the intrinsic scatter in the science field
appropriately.

The slope $\beta_\lambda$ ($=E(K_s-\lambda)/E(H-K_s)$) of the
reddening vector is calculated as:

\begin{equation}
    \label{eq:lines}
    \beta_\lambda = \frac{\mathrm{Cov}(x,y) - \mathrm{Cov}(\epsilon^x, \epsilon^y) -
      \mathrm{Cov}(x^\mathit{cf},y^{cf}) + \mathrm{Cov}(\epsilon^\mathit{cfx},
      \epsilon^\mathit{cfy})}{\mathrm{Var}(x) - \mathrm{Var}(\epsilon^x) -
      \mathrm{Var}(x^\mathit{cf}) + \mathrm{Var}(\epsilon^\mathit{cfx})}
\end{equation}

\noindent where $x$ and $y$ are the $(H-K_s)$ and the $(K_s-\lambda)$
colors, respectively, and $\mathrm{Cov}(x,y)$,
$\mathrm{Cov}(\epsilon^x, \epsilon^y)$, $\mathrm{Var}(x)$, and
$\mathrm{Var}(\epsilon^x)$ are the covariance of the $(x,y)$
distribution, the covariance of the $(x,y)$ errors, the variance of
the $x$ distribution, and the variance of the errors in $x$,
respectively; the symbols marked with $\mathit{cf}$ are the same but
refer to the control field. The uncertainty in $\beta_\lambda$ is
estimated using the bootstrap method corrected by a factor of $1.25$
as described in detail in Paper I.

Figures \ref{fig:ccd_b59} and \ref{fig:ccd_fest} show the best LINES
fits to the color-color diagrams of B59 and FeSt, respectively.  The
fit is naturally weighed more heavily toward the low-extinction end,
since it contains most of the datapoints. The apparent deviation
between the fit line and the data points at high extinction is not
considered to be statistically relevant (see
Sect. \ref{sec:ext-law-vs-density}).

The extinction law $A_\lambda/A_\mathit{K_s}$ is derived from $\beta_\lambda$
as:

\begin{equation}
    \label{eq:ak_alambda}
    \frac{A_\lambda}{A_\mathit{K_s}} = 1-\left(\frac{A_H}{A_\mathit{K_s}}-1\right) \beta_\lambda  \; .
\end{equation}

\noindent The uncertainty in the ratios $A_\lambda/A_\mathit{K_s}$ is
calculated by error propagation from the uncertainty in
$\beta_\lambda$. We assume a ratio $A_H/A_\mathit{K_s}$ of 1.55
following \citet{Indebetouw05} since we cannot determine it accurately
from our data. The choice of this anchor point is crucial for the
comparison between different observations and between observations and
models, since different values will produce different extinction
laws. We chose this value over others in the literature (for example,
most recently, \citet{Nishiyama:2009aa} found
$A_H/A_\mathit{K_s}=1.62$ toward the Galactic Center) because it is
the most widely used in previous studies and therefore allows for the
most direct comparison. Also, using this value, $\beta_J$ translates
into a value of $A_J/A_\mathit{K_s}$ in very close agreement with the
models considered here (see Fig. \ref{fig:ext_law}). All these models
agree on the value of $A_J/A_\mathit{K_s}$ even though they refer to
different grain size distributions, so this agreement lends support to
our assumption. Regardless, our results can be easily adapted to
another definition should it prove to be more appropriate (see also
Sect. \ref{sec:comparison-cores}).

\subsection{The extinction law in B59 and FeSt}
\label{sec:results}

\begin{figure}
\center
  \resizebox{8cm}{!}{\includegraphics{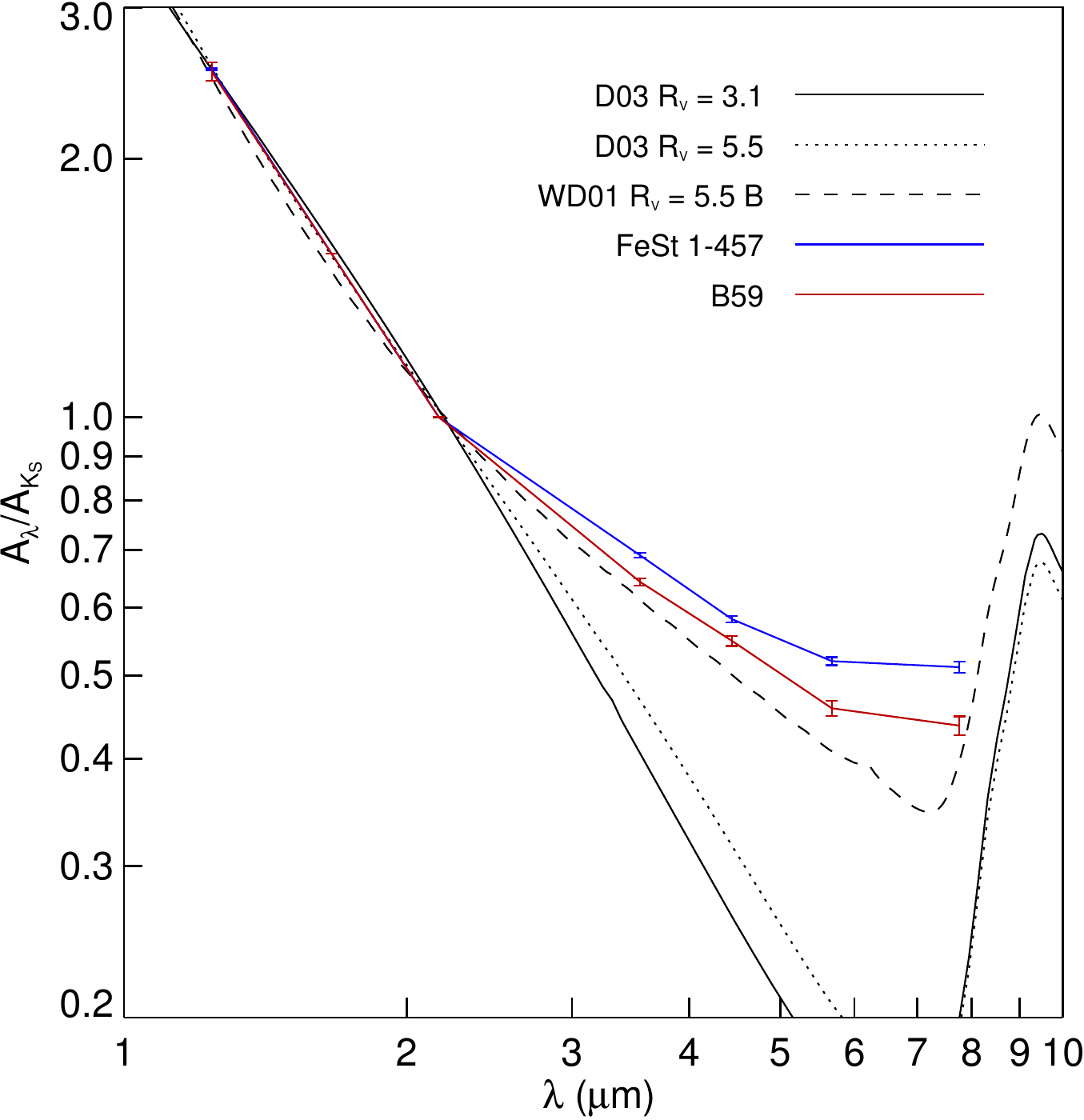}}
\caption{The extinction law for FeSt ({\it blue line}) and B59 ({\it
    red line}). Also shown are the \citet{Draine03b,Draine03c} models
  for $R_V=3.1$ ({\it solid line}) and $R_V=5.5$ ({\it dotted line}),
  and the \citet{Weingartner01} models for $R_V=5.5$, case B ({\it
    dashed line}).}
\label{fig:ext_law}
\end{figure}

\begin{table*}
  \caption{Slopes $E(K_s-\lambda)/E(H-K_s)$ and the extinction law $A_\lambda/A_\mathit{K_s}$ as a
    function of wavelength for B59 and FeSt.}
\label{table:results}
\centering
\onehalfspacing
\begin{tabular}{c c c c c c c c c}
\hline\hline
\multirow{2}{*}{Band}&& \multirow{2}{*}{$\lambda$ ($\mu$m)}&& \multicolumn{2}{c}{B59} & & \multicolumn{2}{c}{FeSt 1-457}\\
 \cline{5-6} \cline{8-9}
& & & & $\beta_\lambda=E(K_s-\lambda)/E(H-K_s)$ & $A_\lambda/A_\mathit{K_s}$ & & $\beta_\lambda=E(K_s-\lambda)/E(H-K_s)$ & $A_\lambda/A_\mathit{K_s}$ \\
\hline
$J$ && 1.240 &&  -3.051 $\pm$ 0.114 & 2.525 $\pm$  0.063 & & -3.081 $\pm$ 0.011 & 2.541  $\pm$  0.006 \\
$H$ && 1.664 &&  --- & 1.550  & & --- & 1.550 \\
$K_s$ && 2.164 &&  --- & 1.000 & & --- & 1.000 \\
$[3.6]$ && 3.545 && 0.649 $\pm$ 0.009  & 0.643 $\pm$ 0.007 & & 0.558 $\pm$ 0.006 & 0.691 $\pm$ 0.004 \\
$[4.5]$ && 4.442 && 0.820 $\pm$ 0.011  & 0.549 $\pm$ 0.007 & & 0.754 $\pm$ 0.008 & 0.582 $\pm$ 0.005 \\
$[5.8]$ && 5.675 && 0.985 $\pm$ 0.013  & 0.458 $\pm$ 0.010 & & 0.866 $\pm$ 0.008 & 0.520 $\pm$ 0.005 \\
$[8.0]$ && 7.760 && 1.023 $\pm$ 0.017  & 0.437 $\pm$ 0.011 & & 0.882 $\pm$ 0.011 & 0.512 $\pm$ 0.008 \\
\hline
\end{tabular}
\end{table*}

The derived slopes $\beta_\lambda$ and the corresponding extinction
law $A_\lambda/A_\mathit{K_s}$ from 1.2 to 8.0 $\mu$m are given in Table
\ref{table:results} for B59 and FeSt, and graphically in Figure
\ref{fig:ext_law} ({\it red} and {\it blue lines} for B59 and FeSt,
respectively). Also shown in the figure are the models from
\citet[][hereafter referred to as WD01]{Weingartner01}: the {\it
  solid} and {\it dotted lines} represent the ``case A'' model of
WD01, re-normalized by
\citet{Draine03b,Draine03c}\footnote{\url{http://www.astro.princeton.edu/~draine/dust/dustmix.html}},
for $R_V$ of 3.1 and 5.5, respectively, and the {\it dashed line}
represents the ``case B'' model of WD01 for $R_V$ of 5.5.

In the near infrared only the $J$ band relative extinction can be
assessed, since $A_\mathit{K_s}/A_\mathit{K_s}=1$ and we fixed $A_H/A_\mathit{K_s}=1.55$
(see Sect. \ref{sec:lines}). Our derived $A_J/A_\mathit{K_s}$ for both B59
and FeSt are in perfect agreement with the dust grain models, and are
also indistinguishable from other measurements in the literature for
this wavelength (see Sect. \ref{sec:comparison-lit}). This agreement
supports our assumption of $A_H/A_\mathit{K_s}=1.55$.
 
In the IRAC bands, from 3.6 to 8.0 $\mu$m, both cores show a very flat
extinction law, in particular flatter than the $R_V=5.5$ ``case B''
model of WD01.

\subsection{The extinction law as a function of column density}
\label{sec:ext-law-vs-density}

\begin{figure*}
\includegraphics[width=\textwidth]{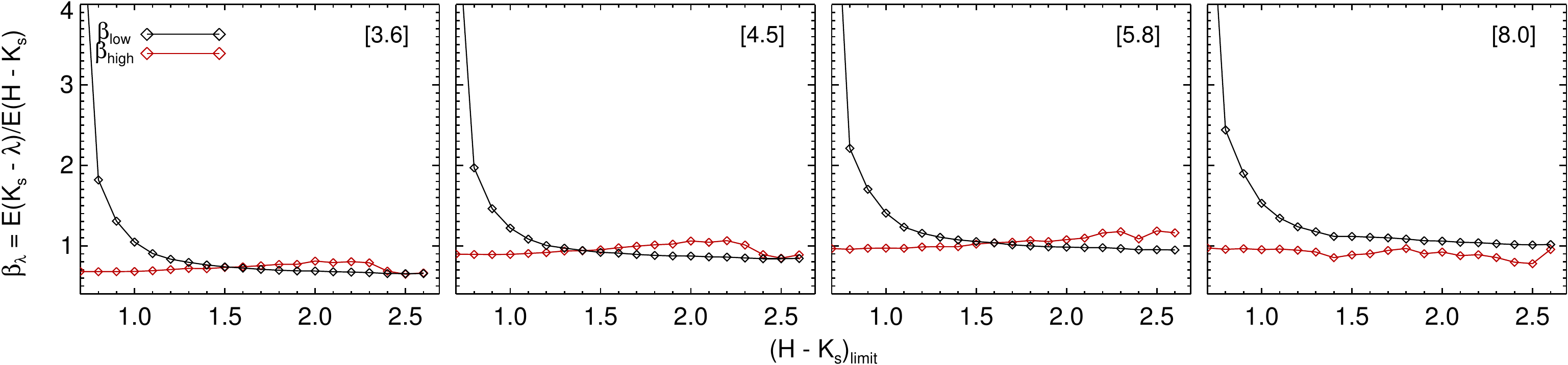}
\caption{$\beta_\mathit{low}$ and $\beta_\mathit{high}$ for B59 as a
  function of $(H-K_s)_\mathit{limit}$ for each IRAC band, [3.6],
  [4.5], [5.8], [8.0] from left to right, showing no evidence for a
  break in the extinction law with column density. The scale in the
  y-axes is the same for all panels.}
\label{fig:2slopes_b59}
\end{figure*}

\begin{figure*}
\includegraphics[width=\textwidth]{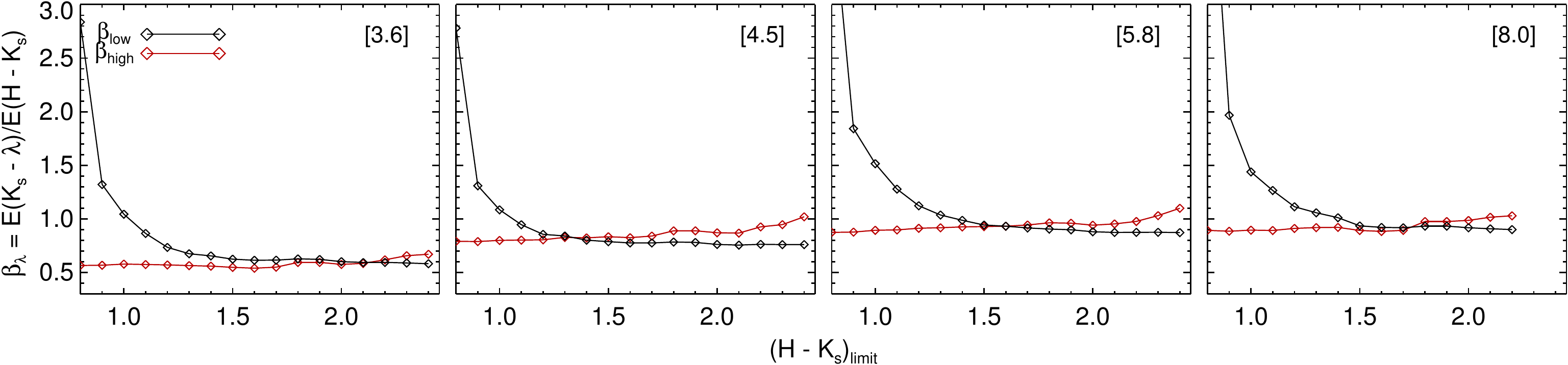}
\caption{$\beta_\mathit{low}$ and $\beta_\mathit{high}$ for FeSt as a
  function of $(H-K_s)_\mathit{limit}$ for each IRAC band, [3.6],
  [4.5], [5.8], [8.0] from left to right, showing no evidence for a
  break in the extinction law with column density. The scale in the
  y-axes is the same for all panels.}
\label{fig:2slopes_fest}
\end{figure*}

We analyzed the behavior of the extinction law as a function of
extinction to see whether we found a change that could be attributed
to grain growth at any specific column density threshold. Figures
\ref{fig:2slopes_b59} and \ref{fig:2slopes_fest} show that the
datapoints at high extinctions depart slightly from the LINES fit,
suggesting a change of behavior at some column density. In Paper I we
showed that the LINES method is robust enough to detect such a
change. Using the ($H-K_s$) color, we divided the sample in
low-extinction ($(H-K)$ less than a value $(H-K)_\mathit{limit}$), and
high-extinction ($(H-K) > (H-K)_\mathit{limit}$) subsamples, and
determined the best fits to the reddening vector in the two groups
using LINES, obtaining two slopes $\beta_\mathit{low}$ and
$\beta_\mathit{high}$ for each $(H-K)_\mathit{limit}$. This was done
for increasing values of $(H-K)_\mathit{limit}$ in steps of 0.1 mag.
Figures \ref{fig:2slopes_b59} and \ref{fig:2slopes_fest} show the
results for B59 and FeSt, respectively. According to the tests with
synthetic data from Paper I, a break in the extinction law would show
as a significant difference between the {\it black} (low extinction)
and the {\it red} (high extinction) curves, which is not
observed. Based on this analysis we conclude that the departure of the
datapoints at high extinction from the LINES fit observed in the
color-color diagrams is not statistically significant. We do not find
any change of the extinction law with column density for either core.

\section{Discussion}
\label{sec:discussion}

\subsection{Comparison with previous work: a universal MIR extinction
  law?}
\label{sec:comparison-lit}

In the near-infrared the extinction law is well described by a
power-law ($A_\lambda \propto \lambda^{-\alpha}$) with indexes tightly
ranging from 1.6 to 1.85. This constancy has led authors to use the
near-infrared as anchor to normalize the extinction law in the MIR.
Although early studies reported that the MIR extinction law continued
as a power-law from the NIR \citep{Whittet88,Martin:1990zr,Landini84},
most recent studies find that it departs from the power-law around 3
$\mu$m and becomes flat (also referred to as gray) at least until the
silicate feature around 9.7 $\mu$m.

\begin{figure*}
\includegraphics[width=\textwidth]{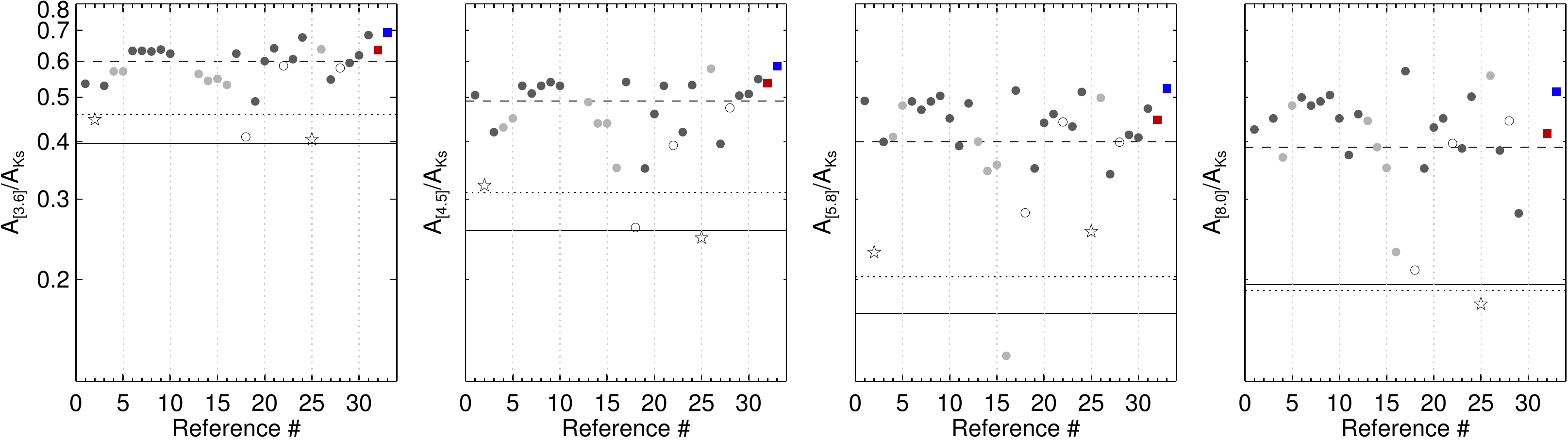}
\caption{The extinction law for FeSt ({\it blue filled squares}) and
  B59 ({\it red filled squares}), in comparison with the extinction
  laws from the literature, shown for each IRAC band separately for
  clarity. The scale in the y-axes is the same for all panels.  From
  left to right, $A_\lambda/A_\mathit{K_s}$ for the IRAC bands [3.6],
  [4.5], [5.8], and [8.0] is shown for each reference in the
  literature. For the works not based on IRAC data the plotted values
  were the closest match in wavelength. The {\it dark circles} refer
  to extinction laws derived toward lines of sight containing dense
  regions, the {\it light circles} toward the diffuse medium, the {\it
    open circles} toward low (column) density lines of sight in the
  direction of star forming regions, and the {\it star symbols} toward
  regions dominated by outflows. Also shown are the
  $A_\lambda/A_\mathit{K_s}$ values from the
  \citet{Draine03b,Draine03c} models for $R_V=3.1$ ({\it solid line})
  and $R_V=5.5$ ({\it dotted line}), and the \citet{Weingartner01}
  models for $R_V=5.5$, case B ({\it dashed line}) for the
  corresponding wavelengths. All extinction laws are normalized to
  $A_H/A_\mathit{K_s} = 1.55$. The references are as follows: \newline
  \small {\bf (1)} \citet{Lutz:1999aa}; {\bf (2)} \citet{Bertoldi99}
  and \citet{Bertoldi99}; {\bf (3-5)} \citet{Indebetouw05} for RCW49;
  $l=284$\degr; $l=42$\degr; {\bf (6-10)} \citet{Flaherty:2007aa} for
  2024/2023; 2068/2071; Serpens; Orion; Ophiuchus; {\bf (11-12)}
  \citet{McClure:2009aa} for $A_\mathit{K_s}<1$ mag;
  $A_\mathit{K_s}>1$ mag; {\bf (13-16)} \citet{Zasowski09} for
  $10\degr<|l|<15\degr$; $30\degr<|l|<40\degr$; $50\degr<|l|<60\degr$;
  $|l|>90\degr$; {\bf (17)} \citet{Nishiyama:2009aa}; {\bf (18-21)}
  \citet{Chapman:2009ab} for $0 < A_\mathit{K_s} \leq 0.5$; $0.5 <
  A_\mathit{K_s} \leq 1$; $1 < A_\mathit{K_s} \leq 2$; $A_\mathit{K_s}
  \geq 2$ mag; {\bf (22-25)} \citet{Chapman:2009aa} for $0 <
  A_\mathit{K_s} \leq 0.5$; $0.5 < A_\mathit{K_s} \leq 1$;
  $A_\mathit{K_s}>1$ mag; regions with outflows; {\bf (26)}
  \citet{Gao09}; {\bf (27)} \citet{Fritz11}; {\bf (28-29)}
  \citet{Olofsson11} for low and high column densities; {\bf (30-31)}
  \citet{Cambresy11} for $A_V<15$ mag; $A_V>20$ mag; {\bf (32-33)}
  This work for B59 and FeSt.}
\label{fig:literature-irac}
\end{figure*}

Figure \ref{fig:literature-irac} summarizes the extinction laws
determined by previous works, and shows that most regions have flat
extinction laws. Each panel shows $A_\lambda/A_\mathit{K_s}$ for each
of the IRAC bands ([3.6], [4.5], [5.8], and [8.0]) as a function of
the corresponding reference in the literature. For the same reference,
a gray or flat extinction law will show approximately the same value
in all panels, whereas an extrapolation of the NIR power-law will show
as a decrease from the left- to the right-hand panels.
\citet{Lutz:1996aa} and \citet[][Ref. \#1]{Lutz:1999aa} first found a
flat MIR extinction law toward the Galactic Center using spectroscopy
of hydrogen recombination lines. \citet{Viehmann05} later measured a
consistently flat extinction at 4.66 $\mu$m toward the same region,
and \citet[][Ref. \#17]{Nishiyama:2009aa} and \citet[][Ref.
\#27]{Fritz11} independently confirmed a flat MIR extinction law
toward that line of sight. Based on {\it Spitzer} photometry,
\citet{Indebetouw05} found a flat MIR extinction law toward other
lines of sight crossing both dense (Ref. \#3) and diffuse (Ref. \#4-5)
media. A number of other studies using {\it Spitzer} followed, namely
\citet[][Ref. \#6-10]{Flaherty:2007aa}, \citet[][Ref.
\#18-21]{Chapman:2009ab} and \citet[][Ref. \#11-12]{McClure:2009aa},
who found gray extinction laws for star forming regions and molecular
clouds, and \citet{Roman-Zuniga07}, \citet[][Ref.
\#22-25]{Chapman:2009aa} and \citet[][Ref. \#28]{Olofsson11}, who
found gray extinction laws for cloud cores. Specifically targeting low
density lines of sight, \citet[][Ref. \#14-16]{Zasowski09} found flat
MIR extinction laws over a very wide portion of the Galactic midplane.
The extinction laws of FeSt ({\it red}, Ref. \#32) and B59 ({\it blue
  filled circles}, Ref. \#33) follow the same flat trend in the IRAC
bands. The column densities probed by our data ($A_V$ up to 50 mag)
are among the highest ever studied in this context, and correspond to
high local volume densities as attested by the presence of
high-density tracer molecules
\citep{Rathborne08,Frau10,Frau12}. Still, despite the difference in
density, the extinction laws of these cores are unremarkable with
respect to less dense environments.

Some authors (\citet[][Ref. \#18-21]{Chapman:2009ab}, \citet[][Ref.
\#11-12]{McClure:2009aa}, \citet[][Ref. \#28-29]{Olofsson11}, and
\citet[][Ref. \#30-31]{Cambresy11}) surveyed lines of sight of
different column densities within single regions and found a slight
flattening trend toward higher densities, but only the first find
significantly steep laws in absolute terms. In contrast, other studies
have looked for but failed to find any significant variation of the
extinction law with increasing density \citep[][Ref
\#22-24]{Chapman:2009aa}, as do we for B59 and FeSt (Sect.
\ref{sec:ext-law-vs-density}), even though we probe a large range in
extinction.

There are only three exceptions in the recent literature where the
observed MIR extinction laws are steep enough to follow the power-law
extrapolation from the NIR: \citet[][Ref. \# 18]{Chapman:2009ab},
toward the low column density ($A_\mathit{K_s}<0.5$ mag) regions of
Perseus, Ophiuchus and Serpens; \citet{Bertoldi99} and
\citet[][Ref. \#2]{Rosenthal00}, toward the outflow region of the
Orion OMC-1; and \citet[][Ref. \#25]{Chapman:2009aa}, also in the
direction of outflows. These differences are interesting and require
further study but represent a small minority of the studies in the
literature.

From the data collected so far, there is no clear indication that the
MIR extinction law varies notably from region to region, or that it
varies as a function of density, since diffuse and dense regions alike
show flat extinction laws. The MIR extinction law does not seem to
depend (uniquely) on density, at least within the (column) density
regimes probed by current observations.

\subsection{Comparison with models: implications for grain sizes}
\label{sec:comp-with-models}

As mentioned briefly in the Introduction, the extinction can be used
to study the grain size distribution of the interstellar dust. The
extinction law in the optical and ultraviolet has been reported to
change from region to region, presumably due to the size distribution
of the grains causing the extinction
\citep[e.g.,][]{Savage79,Whittet03}, with a correlation between larger
grains and higher densities\footnote{``Larger'' grains and
  ``higher''densities refer here to the larger grains that can be
  probed with optical and UV extinction laws ($<0.2~\mu$m) and to the
  largest column densities that can be probed with optical and UV
  observations ($A_V$ of a few magnitudes).}  interpreted as evidence
for grain growth \citep{Cardelli89}. The extinction law at these
wavelengths is however only sensitive to grain sizes less than $\sim
0.2~\mu$m \citep{Draine11}. To probe the size distribution of larger
grains, the infrared extinction law must be used instead.

The observed flat MIR extinction laws are best described by the case
B, $R_V$ of 5.5 model of \citet {Weingartner01}. Case B refers to a
grain size distribution that includes a significant fraction of large
(up to 10 $\mu$m) grains, whereas their case A model includes grains
with maximum sizes of $\sim 1~\mu$m and produces a steep MIR
extinction law\footnote{Cases A and B are indistinguishable from each
  other at optical and UV wavelengths.}. Therefore, flat extinction
laws are believed to be produced by dust containing significant
fractions of large grains. Figure \ref{fig:literature-irac} shows that
the case B, $R_V$ of 5.5 model of WD01 adequately describes the
observed extinction laws at 3.6 $\mu$m and 4.5 $\mu$m but that it does
not produce a gray enough extinction law to match most observations at
5.8 $\mu$m and 8.0 $\mu$m. Since most of these observations are based
on broad-band data, this may indicate that there are narrow,
unresolved emission features contributing to the extinction law at
these wavelengths. Alternatively, it may mean that the grain size
distribution corresponding to these models still does not contain
enough large grains, or large enough grains to explain the
observations.

As mentioned before, grains are believed to grow at the high densities
and low temperatures of shielded molecular clouds, where the formation
of ice mantles and grain-grain collisions are most favored. However,
observations show that the extinction law is not (uniquely) dependent
on density (see Figure \ref{fig:literature-irac} and Sect.
\ref{sec:comparison-lit}), so either other factors determine the shape
of the MIR extinction law, or large grains can form and exist at
relatively low densities. Even though the case A, $R_V=3.1$ model of
WD01 (later renormalized by \citet{Draine03b,Draine03c}) without large
grains has been considered to be the appropriate model for the diffuse
interstellar medium, it is actually hardly ever observed in the MIR.

The extinction laws of B59 and FeSt are also comparable to, and
actually flatter than the case B, $R_V=5.5$ model of WD01 (Figure
\ref{fig:ext_law}), suggesting that these cores contain a significant
fraction of large grains if the models are taken at face value. Unlike
\citet{Cambresy11}, who find a flattening of the extinction law at a
threahold of $A_V=20$ mag, we do not find any such change in the
extinction laws of B59 or FeSt (Sect. \ref{sec:ext-law-vs-density})
even though our data cover a large enough range of extinction and our
method has been shown to be able to detect it (Paper I), further
suggesting that large grains are present already in the low-density
edges of these cores. However, this is somewhat inconsistent with the
results of \citet{Aguti:2007aa}, who find depletion of C$^{18}$O
toward FeSt at relatively low extinctions ($A_V=12$ mag). If the
accretion of ice mantles by grains as molecular species condense onto
their surfaces increases their sizes and promotes their growth, the
depletion of molecules should be regarded as indirect evidence for
grain growth. For completeness, \citet{Frau12} suggest that there is
no depletion in B59, although their arguments are solely based on a
high abundance of CS\footnote{Unpublished radio data of B59 suggest
  that
  there may be depletion of C$^{18}$O at column densities higher than
  $A_V=30$ mag (Rom\'an-Z\'u\~niga, private communication).}.

It is clear from this study and from the literature that we still do
not have a consistent understanding of the extinction law and/or of
the dust grain properties in the different environments.

\subsection{Comparison between the two cores}
\label{sec:comparison-cores}

In detail, the MIR extinction laws of FeSt and B59 are not identical:
the extinction law of FeSt is grayer than that of B59 at the $5$- to
$12$-$\sigma$ level depending on the wavelength. This is true as long
as our assumption that the NIR extinction law is the same for the two
lines of sight applies. When converting the reddening laws (the color
excess ratios $\beta_\lambda$, Sect. \ref{sec:lines}) to extinction
laws ($A_\lambda/A_\mathit{K_s}$), we assumed $A_H/A_\mathit{K_s}=1.55$
following \citet{Indebetouw05}, which corresponds to a NIR extinction
law being a power law with index 1.65 ($A_\lambda \propto
\lambda^{-{1.65}}$). This value is adopted by many studies of the
extinction law as universal but in fact different values have been
measured toward different lines of sight. The values in the literature
range between 1.85 and 1.65 (see references in the Introduction), and
there is no understanding of how it changes from region to region. It
is therefore possible that B59 and FeSt have different extinction laws
in the NIR. In particular, if the index of the NIR extinction law for
FeSt is the adopted 1.65, but that for B59 is instead 1.90, the two
extinction laws are indistinguishable within the errors. However, as
mentioned in Sect. \ref{sec:results}, we measure very similar
$A_J/A_\mathit{K_s}$ values for the two cores, supporting the
assumption of similar NIR extinction laws for the two cores. In the
following we take this assumption as correct and discuss the
implications of the MIR extinction laws for the two cores being
different, as shown in Figure \ref{fig:ext_law}.

We also assume, for the sake of argument and based on the models of
WD01, \citet{Draine03b} and \citet{Draine03c}, that the flatter the
extinction law, the larger the grains causing the extinction. In this
context, FeSt should have larger grains (or a larger fraction of large
grains) than B59, since it has the flattest extinction law. The
depletion of C$^{18}$O at relatively low extinctions observed in FeSt
in contrast to the lack of depletion in B59 further suports the
presence of larger grains in FeSt.

Why are the grains smaller in B59 than in FeSt? Since the two cores
belong to the same cloud it is reasonable to assume that they have the
same chemical composition, so we do not expect that different
metallicities are the source of the different extinction laws. Our
data probe approximately the same range of column densities for both
lines of sight, and both FeSt and B59 are dense cores, so the column
density is likely to represent approximately the same physical
density. If this is the case, then the difference in grain size
distribution for the two cores cannot be explained by density
arguments either. If density is not the dominant factor, it may be
that the different internal conditions of FeSt and B59 are causing the
difference in extinction laws. In fact, the most obvious difference
between the two cores is that FeSt is quiescent, whereas B59 contains
$\sim 20$ young stellar objects \citep{Brooke07,Forbrich09,Covey10}
and at least two molecular outflows
\citep{Onishi99,Riaz09,Roman-Zuniga10,Duarte-Cabral12}. The young
stars necessarily increase the temperature of the core, at least
locally, and their winds and outflows inject mechanical energy and
turbulence into the intra-core medium. It would therefore not be
unexpected that the newly formed stars have some bearing in the
reprocessing of the dust grains. If star formation is instrumental in
the destruction of large grains, then the extinction law in B59 would
be steeper than that of FeSt, as is observed. This would be consistent
with the other observation of steep extinction laws toward other, more
exposed outflow regions \citep{Bertoldi99,Rosenthal00,Chapman:2009aa}.

We note that \citet{Alves:2008aa} find a magnetic field strength
toward FeSt around 65 $\mu$G, almost 4 times higher than that toward
B59, corresponding to degrees of polarization reaching 15\% in the
region of FeSt and around 1\% for B59. However, the strength of the
magnetic field is related to the alignment of the grains, and
therefore is not very revealing to the understanding of the grain size
distribution. Also, since this study was done with R-band
observations, it characterizes the magnetic field of the lower-density
regions ($A_V<5$ mag) of the cloud, so the ambient magnetic field
instead of that of the cores

\section{Conclusions}
\label{sec:conclusions}

We applied the unbiased LINES method \citep{Ascenso12} to determine
the extinction law from photometric data from 1.2 to 8.0 $\mu$m toward
the two densest cores in the Pipe Nebula, B59 and FeSt 1-457.

We find a flat or gray extinction law for both cores. The comparison
with models suggests that the dust in these cores contains a
significant fraction of large grains. Despite the wide range of column
density probed in this study (up to 50 magnitudes of visual
extinction) with respect to most previous studies, the extinction laws
we derive are not remarkably different from others toward other lines
of sight. In particular a review of the literature indicates that the
mid-infrared extinction law is gray already at low column densities,
suggesting that either large grains are present already in low column
density regions and in the diffuse medium, or that the existing dust
models need to be revised at mid-infrared wavelengths. The
implications for the theories of grain growth, that generally require
high densities for the formation of large grains, are not discussed in
this paper but should be addressed appropriately in the future.

Although the extinction laws are gray for the two cores, that of FeSt
is grayer than that of B59. Taken at face value this suggests that
FeSt has larger grains, or a larger fraction of large grains, than
does B59. We hypothesize that this reflect the internal differences in
the two cores, in particular with the presence of young stars and
outflows in B59 and lack thereof in FeSt. We propose that the grains
have already started to be reprocessed in B59 by the increase in
temperature and turbulence caused by the young stars.

We find no evidence for a flattening of the extinction law as a
function of column density, and conclude that the low-density ($A_V
\lesssim 10$ mag) edges of the cores already contain large grains.

\begin{acknowledgements}
  The authors thank D. Lutz and N. Chapman for kindly providing the
  electronic versions of their extinction laws. The research leading
  to these results has received funding from the European Community’s
  Seventh Framework Programme (/FP7/2007-2013/) under grant agreement
  No 229517. Support for this work was also provided by NASA through
  an award issued by JPL/Caltech, contract 1279166. CRZ acknowledges
  support from Program CONACYT 152160, M\'exico.
\end{acknowledgements}

\bibliographystyle{aa}
\bibliography{/Users/jascenso/Dropbox/Science/bib}

\end{document}